\documentclass[pre,aps,floatfix,twocolumn,showpacs,preprintnumbers,superscriptaddress]{revtex4}
\usepackage{graphicx}
\usepackage{dcolumn}
\usepackage{bm}
\usepackage{amssymb}
\usepackage{pifont}
\usepackage{amsmath}
\usepackage{times}

\begin{document}

\title{Universal scaling of Lyapunov-exponent fluctuations in space-time chaos}

\author{Diego Paz\'o} 
\author{Juan M.\ L{\'o}pez}
\affiliation{Instituto de F\'{\i}sica de Cantabria (IFCA), 
CSIC--Universidad de Cantabria, E-39005 Santander, Spain}
\author{Antonio Politi}
\affiliation{Institute for Complex Systems and Mathematical Biology, King's
College, University of Aberdeen, Aberdeen AB24 3UE, United Kingdom}
\date{\today}

\begin{abstract}
Finite-time Lyapunov exponents of generic chaotic dynamical systems fluctuate in
time. These fluctuations are due to the different degree of stability across the
accessible phase-space. A recent numerical study of spatially-extended systems
has revealed that the diffusion coefficient $D$ of the Lyapunov exponents (LEs)
exhibits a non-trivial scaling behavior, $D(L) \sim L^{-\gamma}$, with the
system size $L$. Here, we show that the wandering exponent $\gamma$ can be
expressed in terms of the roughening exponents associated with the corresponding
``Lyapunov-surface". Our theoretical predictions are supported by the numerical
analysis of several spatially-extended systems. In particular, we find that the
wandering exponent of the first LE is universal: in view of the known
relationship with the Kardar-Parisi-Zhang equation, $\gamma$ can be expressed
in terms of known critical exponents. Furthermore, our simulations reveal that
the bulk of the spectrum exhibits a clearly different behavior and suggest that
it belongs to a possibly unique universality class, which has, however, yet to
be identified.
\end{abstract}


\maketitle

\section{Introduction} 
Lyapunov exponents (LEs) $\langle \lambda_i\rangle$ are the most powerful tool
for a detailed characterization of chaotic dynamics~\cite{ott}: they allow
determining the number of unstable directions, the Kolmogorov-Sinai entropy, the
fractal dimension and other dynamical invariants. The number of LEs equals the
number of degrees of freedom $N$. The LEs, ordered from the largest to the
smallest one, define the so-called spectrum of LEs. In spatially extended
systems, $N \propto L^d$ (where $L$ is the system linear size and $d$ is the
dimensionality of the space); in the thermodynamic limit ($L\to\infty$) the
Lyapunov spectrum converges to an asymptotic curve, that is specific to each
system and depends only on $\rho=i/L^d$. The existence of such a limit shape is
taken as the proof of extensivity, i.e. that quantities like the fractal
dimension or the Kolmogorov--Sinai entropy are proportional to the system
volume.

The LEs quantify the exponential expansion (contraction) growth rates along
the $N$ covariant or characteristic directions in the infinite-time limit. In
fact, tangent space can be decomposed into covariant subspaces and the
corresponding vector-base provides relevant information that is encoded in the
so-called characteristic/covariant Lyapunov vectors
(LVs)~\cite{ruelle79,eckmann,legras96}. These vectors actually contribute to
identify the local structure of the invariant measure, to uncover the possible
presence of collective phenomena~\cite{takeuchi09}, to determine the dimension
of the inertial manifold~\cite{yang09}, or to detect spurious LEs observed in
embedded time series~\cite{yang12}, to cite a few applications.

In a time interval $t$, an infinitesimal perturbation pointing along the $i$-th
LV is expanded/contracted in tangent space by a factor $e^{\Lambda_i(t)}$, where
$\Lambda_i(t)$ is the so-called expansion rate. As a result of the variable
degree of instability in phase-space, $\Lambda_i(t)$ fluctuates along the
trajectory (or, equivalently, across phase-space). Nevertheless, the finite-time
Lyapunov exponent (FTLE) $\lambda_i(t) = \Lambda_i(t)/t$ converges, in the
infinite-time limit, to the $i$-th LE, $\lim_{t\to\infty} \lambda_i(t) =
\langle\lambda_i\rangle$, where brackets indicate an average over
trajectories~\cite{foot1} (hereafter, phase-space average and time average are
assumed to coincide). 

It is natural to ask how the FTLE fluctuations scale with both time and system
size. As already argued in \cite{kuptsov11}, this question is not only connected
with the convergence to the thermodynamic limit, but also with the extensivity
of space-time chaos. The best way to approach the problem is by introducing the
(time-dependent) variances,
\begin{equation}
\chi^2_{ij} = \langle{\Lambda_i(t)\Lambda_j(t)}\rangle -
\langle \lambda_i\rangle \langle\lambda_j\rangle t^2 \, ,
\label{chiij}
\end{equation}
and the corresponding diffusion coefficients,
\begin{equation}
D_{ij}= \lim_{t\to\infty} \frac{\chi^2_{ij}}{t} \, .
\label{dij}
\end{equation}
It is expected that for large-enough times, the distribution ${\cal
P}({\bm{\lambda}},t)$ of FTLEs [$\bm{\lambda}(t) = (\lambda_1(t), \dots,
\lambda_N(t))$] is described by a suitable large-deviation function, ${\cal
P}({\bm{\lambda}},t) \propto \exp[-S({\bm{\lambda}})t]$, and
$S(\langle\bm{\lambda} \rangle)=0$. Under the fairly general assumption that the
LEs fluctuations are short-range in time, (this is typically true away from
bifurcations and phase transitions), the central limit theorem implies that
$S(\bm{\lambda})$ has a quadratic structure around its minimum, i.e. that ${\cal
P}({\bm{\lambda}},t)$ is Gaussian,
\begin{equation}
{\cal P}({\bm{\lambda}},t) \propto
\exp\left[-\frac{t}{2} ({\bm{\lambda}}-\langle{\bm{\lambda}}\rangle)
\mathrm{\mathbf{Q}}
({\bm{\lambda}}-\langle{\bm{\lambda}}\rangle)^\dag\right] \, ,
\label{gaussian}
\end{equation}
where $\dag$ denotes the transpose, while the matrix $\mathbf{Q}$ is the inverse
of the symmetric diffusion matrix, {\it i.e.}, $\mathbf{Q}=\mathbf{D}^{-1}$.
Thus, within the Gaussian approximation, $\mathbf{D}$ describes how strong FTLE
fluctuations are for a given system size. In particular, the diagonal elements
$D_{ii}$ correspond to the diffusion coefficient of the expansion rates
$\Lambda_i$ around the average growth $\langle\lambda_i\rangle t$.

Chaos extensivity would naively suggest that the diffusion coefficients should
scale as $L^{-d}$ with the system size. Based on the numerical simulation of a
variety of systems in $d=1$ Kuptsov and Politi~\cite{kuptsov11}, have however
found that $D_{ii}(L) \sim L^{-\gamma}$ with a wandering exponent that is
smaller than 1: $\gamma\simeq 0.85$ for $i>1$, while $\gamma = 1/2$ for $i=1$.
At the same time, it was found that the off-diagonal terms decay as $L^{-1}$ and
so do the eigenvalues of the matrix $\mathbf{D}$, in agreement with the
expectation for an extensive chaotic dynamics. It is rather intriguing that
although the matrix $\mathbf{D}$ becomes increasingly diagonal in the
thermodynamic limit, its diagonal terms scale differently from the matrix
eigenvalues~\cite{kuptsov11}. In the absence of theoretical arguments, one
cannot a priori exclude that the scaling behavior 
of the bulk of the spectrum
is affected by strong
finite-size corrections, so that $\gamma \to 1$ as $L\to\infty$.
We shed some light on this problem
by resorting to the well-known connection
between LV dynamics and the kinetics of rough surfaces. 
This allows unveiling a
theoretical connection between the scaling properties of the FTLEs fluctuations
and the velocity fluctuations of rough surfaces subject to stochastic forces.
This mapping leads to a scaling relation between the universal 
roughening exponents and the FTLE wandering exponent $\gamma$, which thus
turns out to be a true critical exponent.


More specifically, in Section II, we develop the theoretical scaling arguments
which lead us to derive the relevant mathematical relationships. In Section III,
our systematic investigation of the maximal Lyapunov exponent confirms the
theoretical expectations as well as the relationship with Kardar-Parisi-Zhang
(KPZ) dynamics~\cite{kpz}. Section IV is devoted to the entirely new analysis of
the bulk of the spectrum which suggests the correspondence with some yet unknown
stochastic field theory. Finally, in Section V, we summarize the main results
and briefly discuss the open problems.

\section{Theory}
In this section we focus on the scaling behavior of the diagonal elements
$D_{ii}$ of the matrix $\mathbf{D}$. For the sake of simplicity, from now on, we
drop the index $i$ in the formulas and reintroduce it in the next sections, when
it will be necessary to distinguish between different values of the integrated
density $\rho=i/L^d$.

In the following we show that $D(L) \sim L^{-\gamma}$, where the exponent
$\gamma$ can be expressed in terms of the scaling properties of the
corresponding LV, $v(\bm x, t)$. Our scaling theory is based upon the well-known
interpretation of the dynamics of a LV as the statistical evolution of a rough
surface~\cite{pik94,pik98,pik01,szendro07,pazo08}.

For each given LV $v({\bm x},t)$ we define an associated ``surface'' field
$h({\bm x},t)$ through the logarithmic transformation $h({\bm x},t) = \ln
|v({\bm x},t)|$. The LV surface so defined is known to be generically rough and
scale-invariant~\cite{pik94,pik98,pik01,szendro07,pazo08}. As we are interested
in the expansion factor of the LV over a time interval $t$, it is useful to
introduce the field
\begin{equation}
\phi({\bm x},t)=h({\bm x},t)-h({\bm x},0) \, , 
\label{phi}
\end{equation}
which corresponds to the logarithm of the finite-time expansion factor (over a
time $t$). The fluctuations around the average surface position at any given
time are quantified by the surface width $W$,
\begin{equation}
W^2(t,L)= \left \langle \overline{\phi^2({\bm x},t)}\right\rangle - 
\left\langle\overline{\phi({\bm x},t)}^2 \right\rangle \, ,
\label{w2}
\end{equation}
where the angular brackets denote an average over an ensemble of different
trajectories, while the overline is a spatial average (here and in the
following). Scale invariance generically leads to finite-size scaling of surface
fluctuations that can be cast in the typical scaling form~\cite{Barabasi}
\begin{equation}
W^2(t,L) =  L^{2\alpha} \mathcal{F}(t/L^{z}) 
\end{equation}
where $\mathcal{F}(u)$ is a dynamical scaling function, which reaches
asymptotically ($u\to\infty$) a constant value and grows as $u^{2\alpha/z}$ for
small values of $u$. All of the above means that at short times $W^2$ grows as
$t^{2\alpha/z}$, until $t \sim L^z$ when $W^2$ saturates to a size-dependent
value $L^{2\alpha}$. The roughness exponent $\alpha$ and the dynamic exponent
$z$ quantify space-time correlations and fully characterize the statistical and
dynamical behavior of the surface (equivalently, the LV.)

\subsection{Main result}
Before proceeding with the details of the theoretical derivation, we anticipate
our main result, {\it i.e.} that for any FTLE its variance $\chi^2(t,L)$
[defined by \eqref{chiij} with $i=j$] scales as
\begin{equation}
\chi^2(t,L) =  L^{2\alpha-z} \mathcal{G}(t/L^z) t  \, .
\label{eq:scal}
\end{equation}
where $\mathcal{G}(u\gg1)\simeq \mathrm{const}$.
By then comparing this formula with Eq.~\eqref{dij}, this implies that the
LE diffusion coefficient scales as 
\begin{equation}
D(L) =  L^{2\alpha-z} \mathcal{G}(\infty) \, ,
\label{d}
\end{equation}
so that the wandering exponent $\gamma$ heuristically observed
in Ref.~\cite{kuptsov11} is a truly critical exponent, connected to 
the LV surface roughening exponents,
\begin{equation}
\label{main}
\gamma = z-2\alpha \, .
\end{equation}
As a result, $\gamma$ can be determined, once the roughening exponents
of the corresponding surfaces are known. This is valid for any spatial
dimension $d$.
 
\subsection{Derivation of the scaling function $\cal G$}
We now derive Eq.~\eqref{d} and give further details on the scaling of the
diagonal elements of $\mathbf{D}$ in the intermediate regime before saturation.
First, let us notice that the expansion rate $\Lambda(t)$ necessarily refers to
some norm in tangent space. The norm selection is irrelevant for the calculation
of dynamical invariant quantities like the LEs or the diffusion coefficient
$\mathbf{D}$, as they involve an infinite-time limit. However, the finite-time
expansion rates $\Lambda(t)$ depend explicitly on the norm.  Given a
perturbation $v({\bm x},t)$, a rather broad family of $q$-norms
can be defined as follows
\begin{equation}
\label{qnorm}
\|v({\bm x},t) \|_q \equiv \left[ (1/L)^d \sum_{\bm x} \left|v({\bm x},t)
\right|^{q}\right]^{1/q}.
\end{equation}
The standard Euclidean norm corresponds to $q=2$. In the following, our
theoretical arguments will be developed with reference to the 0-norm: $\|v({\bm
x},t) \|_0 = \prod_{\bm x} \left|v({\bm x},t) \right|^{1/L^{d}}$, unless
otherwise specified. The 0-norm is the most convenient and natural choice
because, in this framework, computing the LV norm corresponds to determining the
average height of the surface $h({\bm x},t) = \ln |v({\bm x},t)|$, while the
FTLE corresponds to the surface velocity, so that
\begin{equation}
\Lambda(t)  = \overline{\phi({\bm x},t)} \, ,
\label{eq:Lambda2}
\end{equation}
which, by definition
(\ref{phi}), corresponds to the net displacement of the
average surface position in a time interval $t$.

Now, by combining Eqs.~(\ref{chiij}) and (\ref{eq:Lambda2}) one obtains
\begin{equation}
\chi^2(t,L) = \left\langle \overline{\phi({\bm x},t)}^2 \right\rangle - 
\left\langle \overline{\phi({\bm x},t)}\right\rangle^2 \, .
\label{chi}
\end{equation}
This expression, like Eq.~\eqref{w2} for $W^2$, is a quadratic correlation
function of $\phi({\bm x},t)$ and, therefore, it should scale as
\begin{equation}
\chi^2(t,L) =  L^{2\alpha} \mathcal{G}_\chi(t/L^{z})  \, .
\end{equation}
Consistency with Eq.~\eqref{dij} requires that $\mathcal{G}_\chi(u)$ diverges
linearly with time. This information can be included in the above equation, by
writing the scaling function as a product, $\mathcal{G}_\chi(u) = \mathcal{G}(u)
u$, where $\mathcal{G}$ saturates for $u\to\infty$. Altogether, this leads to
Eq.~\eqref{eq:scal} and the main relation~\eqref{main} for the scaling behavior
of the asymptotic diffusion coefficient.

Although our theoretical arguments are based on the use of the 0-norm, this does
not affect the validity of our main scaling relation~\eqref{main} as, in the
infinite-time limit, the diffusion coefficient $D$ is actually a dynamical
invariant, independent of the norm used.

Before saturation sets in ($t \ll L^z$), the effective diffusion coefficient is
time-dependent and exhibits long-range temporal correlations. These correlations
are important, since they correspond to an anomalous diffusion of the expansion
rate $\Lambda$, as it can be inferred by looking at the ``short-time" behavior
of the dynamic scaling function $\mathcal{G}$ in Eq.~\eqref{eq:scal}, {\it i.e.}
for $1 \ll t\ll L^z$. We indeed find that, still in the 0-norm framework, 
\begin{equation}
\mathcal{G}(u) \approx u^{\nu}  \,\, \mbox{with} \, \, \nu = \frac{2\alpha -z
+d}{z} \, .
\label{eq:delta}
\end{equation}
Let us now show how this result arises from a simple scaling argument. At short
times, correlations only extend over a linear distance of order $\sim t^{1/z}$
so that the system can be considered as formed by a number $N_b \sim
(L/t^{1/z})^d$ of statistically independent blocks. Accordingly, we can write
$\phi({\bm x},t)=\left\langle\, \overline{\phi}\, \right\rangle + \delta
\phi({\bm x},t)$, where $\delta\phi({\bm x},t)$ is the local (intra-block)
fluctuation. From Eq.~\eqref{chi}, we can now estimate $\chi^2$ 
\begin{equation}
 \chi^2 =\left\langle \overline{\phi({\bm x},t) - \left\langle\,
\overline{\phi}\, \right\rangle}^2 \right\rangle =
\left\langle\overline{\delta\phi({\bm x},t)}^2\right\rangle \sim W^2/
N_b\, ,
\end{equation}
so that
\begin{equation}
\chi^2 \sim  \left(\frac{t^{1/z}}{L}\right)^d W^2 \,
\end{equation}
for $t \ll L^z$. By then recalling that $W^2(t) \sim t^{2\alpha/z}$ in this time
regime, we finally obtain Eq.~\eqref{eq:delta}. This concludes our scaling
analysis. 

It is worth to remark that, contrary to the asymptotic behavior,
Eq.~\eqref{eq:delta} is valid only with reference to the 0-norm. This is
because, while the stationary diffusion coefficient implies an infinite-time
limit, the time-dependent effective diffusion coefficient is defined for $t <
L^z$. This will become evident later on from the comparison with numerical
calculations with both the 0-norm and the Euclidean norm.

Let us finally mention that the scaling function Eq.~\eqref{eq:delta} arises
also in the context of the nonequilibrium roughening, where it describes the
velocity-fluctuations~\cite{lopez10,krug91} of a driven interface.

\subsection{Universality}
It is well known that for a wide class of extended dynamical systems, which
include coupled-map lattices, Lorenz-96 model, Kuramoto-Sivashinsky equation,
and many others~\cite{pik98}--- the LV surface $h({\bm x},t)$ associated with
the first LV belongs to the universality class of KPZ~\cite{kpz}. The
universality class of KPZ in extended dynamical systems is, therefore, very
large and includes models which, in spite of relevant differences in the
microscopic details, share a universal behavior of the first LV. This
universality class seems to include all dissipative models with short-range
interactions as well as some symplectic models. Therefore, we expect $D_{11}$ to
be characterized by the same wandering exponent, for all models in the KPZ
universality class. For instance, in $d=1$ one can plug the exact values $\alpha
=1/2$ and $z=3/2$ in Eq.~\eqref{main} to obtain $\gamma = 1/2$.

Regarding the wandering exponent in the bulk LEs ($i\gg1$), it is known that the
corresponding LV surfaces are characterized by a different set of scaling
exponents. This issue has been much investigated in the last few years.
Numerical simulations in $d=1$ suggest that the dynamic exponent is
$z\simeq1$~\cite{szendro07}, and this result is also supported by theoretical
arguments~\cite{pazo08}. Moreover, $2\alpha$ has been found to be much smaller
than that in KPZ, lying in the range 0.15-0.2~\cite{szendro07,pazo08}. Our main
relation in Eq.~\eqref{main} suggests a wandering exponent $\gamma=0.8$-$0.85$
for the diffusion of the bulk LEs that is consistent with the numerical
observations in Ref.~\cite{kuptsov11}.

\section{The Largest Lyapunov exponent: Numerical results}
In this and the next section we compare our theoretical predictions with
detailed numerical simulations of several systems. Here we focus on the
diffusion
coefficient of the first LE.
\begin{figure}
 \centerline{\includegraphics*[width=80mm]{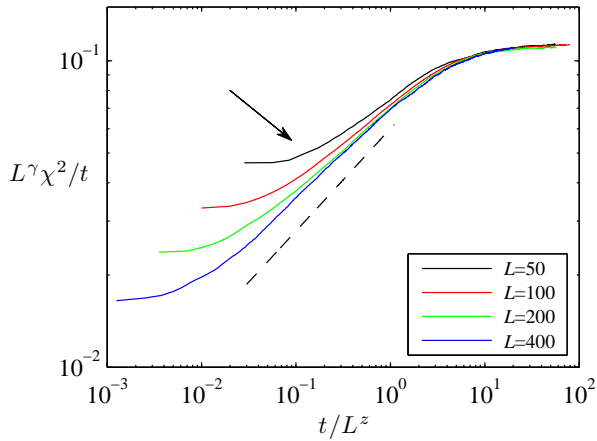}}
\caption{(Color online) Rescaled variance $\chi^2$ of the fluctuations of the maximum FTLE in
a chain of H\'enon maps (see the text for the parameter values). The various
curves correspond to $L=50$, 100, 200, and 400. The exponent values used for the
data collapse are $z=3/2$, $\gamma=1/2$; while the slope of the straight line is
1/3.}
\label{fig1}
\end{figure}

\subsection{One-dimensional systems ($d=1$)}
The first model we analyze is a chain of H\'enon maps, whose LE
fluctuations have been recently studied in Ref.~\cite{kuptsov11}. The model
writes
\begin{equation}
x_n(t+1)=a-[x_n(t)+\epsilon {\cal D} x_n(t)]^2+b x_n(t-1)
\label{henon}
\end{equation}
where ${\cal D}x_n\equiv x_{n-1}-2x_n+x_{n+1}$ is the discrete Laplacian
operator, $n=1,\ldots,L$, and we select $a=1.4$, $b=0.3$, $\epsilon=0.025$. In
Fig.~\ref{fig1} we plot $L^\gamma\chi^2/t$ , obtaining that for different system
sizes the data indeed collapse for $\gamma = 1/2$ onto a dynamic scaling
function $\mathcal{G}$ that follows Eq.~\eqref{eq:scal} and the predicted
asymptotes both above ($t\gg L^z$) and below ($1\ll t \ll L^z$) the crossover
time. In particular, the very good data collapse observed at long times
validates Eq.~\eqref{main}. $\gamma=1/2$ was also observed in
Ref.~\cite{kuptsov11} where the Euclidean norm was used to measure vector
metrics in tangent space, instead of the 0-norm considered in this paper.
This confirms that the norm choice does not affect the stationary behavior. At
shorter times, finite-size corrections are more sizable, but one can
nevertheless appreciate an increasing quality of the data collapse  with $L$.
The initial slope increases with $L$ and approaches the theoretical prediction
$\nu=1/3$. Notice that this means that $\chi^2$ initially grows as $t^{4/3}$,
{\it i.e.}~the Lyapunov dynamics is superdiffusive in the intermediate regime
before saturation of fluctuations.

\begin{figure}
 \centerline{\includegraphics*[width=80mm]{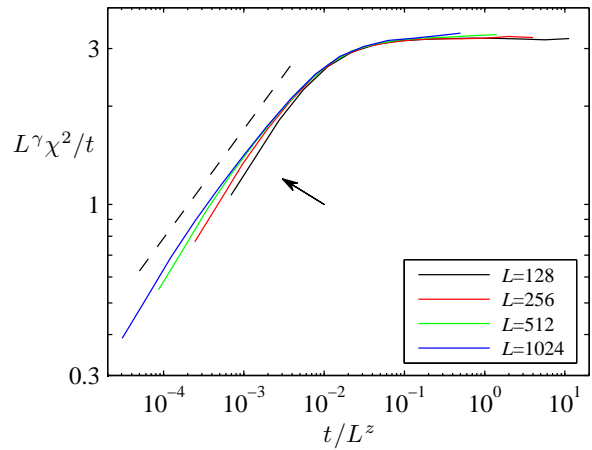}}
\caption{(Color online) Rescaled variance $\chi^2$ in the Lorenz-96 model, Eq.~\eqref{l96},
with $F=8$. The slope of the straight line is  $1/3$.}
\label{fig_l96}
\end{figure}

We have also studied numerically the Lorenz-96 model~\cite{lorenz96} 
\begin{equation} 
\frac{d y_i}{dt} = -y_i-y_{i-1} (y_{i-2}-y_{i+1}) + F \, ,
\label{l96} 
\end{equation}
a time-continuous toy-model of the atmosphere that represents the value of a
scalar variable on a mid-latitude. The data collapse shown in Fig.~\ref{fig_l96}
confirms that the diffusion of the largest LE is well described by
Eq.~\eqref{eq:scal}. The only difference with respect to the previous model is
that, as the arrow indicates, the curves converge towards the asymptotic shape
from below.

\begin{figure}
 \centerline{\includegraphics*[width=80mm]{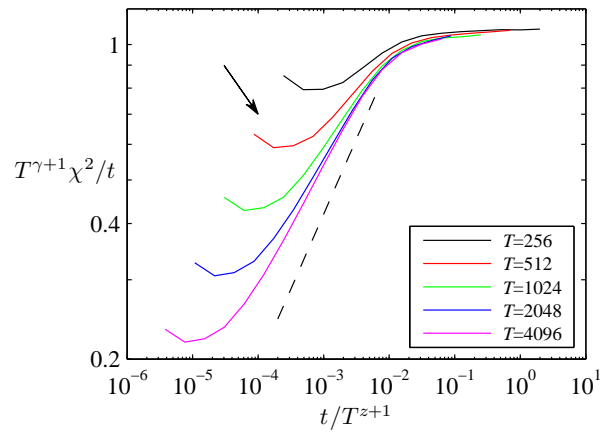}}
\caption{(Color online) Rescaled variance $\chi^2$ in the Mackey-Glass model, Eq.~\eqref{dds},
with $a=0.1$ and $b=0.2$ (like in~\cite{pazo_lopez10}). The slope of the
straight line is  $1/3$.}
\label{fig_mg}
\end{figure}

As a last example of a (pseudo) 1d system, we present our numerical results for
a very different type of chaotic system: a model with delayed feedback. For many
models of this type, the main LV scales as in typical 1d spatio-temporal chaotic
systems~\cite{pik98,pazo_lopez10}, after identifying the delay $T$ with the
system size $L$. Note that one also must rescale the time axis $t$ by a factor
$1/T$, as it is so for the Lyapunov spectrum~\cite{farmer82}. More
specifically, we have carried out simulations of the Mackey-Glass
model~\cite{mg77},
\begin{equation}
\frac{dy(t)}{dt}= -ay(t)+b \frac{y(t-T)}{1+y(t-T)^{10}} \, .
\label{dds}
\end{equation}
The results are plotted in Fig.~\ref{fig_mg}. Apart from a relatively slow
convergence they are again in agreement with our theoretical
predictions.

\subsection{Two-dimensional systems ($d=2$)}
It is very instructive to check the theoretical predictions in two-dimensional
models since the KPZ critical exponents depend on the spatial
dimension. Current numerical capabilities allow us to study a two-dimensional
lattice composed by $L^2$ coupled logistic maps (on a torus geometry)
\begin{equation}
\label{cml}
u_n(t+1)= (1-\epsilon)f[u_n(t)] + \epsilon \sum_{m\in {\cal N}(n)}
f[u_{m}(t)]
\end{equation}
where $f[u(t)] \equiv 4u(t)[1-u(t)]$ and the sum is over the set ${\cal N}$ of
nearest neighbors. 

\begin{figure}
 \centerline{\includegraphics*[width=80mm]{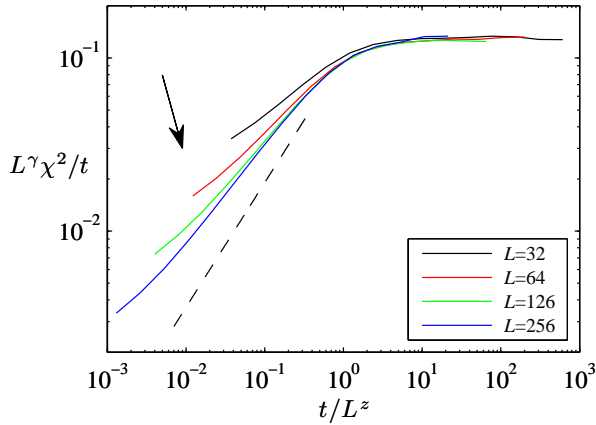}}
\caption{(Color online) Rescaled variance $\chi^2$ of the fluctuations of the maximum FTLE for
a 2d lattice of logistic maps with $\epsilon=0.1$ (the different curves
correspond to $L=32$, 64, 128, and 256 --- from top to bottom. The exponent
values used for the data collapse are $z=1.613$ and $\gamma = 0.839$; while the
slope of the straight line is 0.72.}
\label{fig_2d}
\end{figure}

In two dimensions, only numerical estimates of the KPZ scaling exponents
$\alpha$ and $z$ are available. The best estimations~\cite{forrest90} are
$\alpha\simeq 0.387$, and $z\simeq 1.613$, so that we predict $\gamma\simeq
0.839$ and $\nu \simeq 0.720$ for the wandering exponent in~\eqref{main} and the
time exponent in~\eqref{eq:delta}, respectively. Numerical results for the
dynamic scaling function are plotted in Fig.~\ref{fig_2d}. The data collapse is
excellent, confirming again the validity of our theoretical arguments. Notice
that the LE fluctuations decay with the system size faster in 2d than in 1d 
since $\gamma$ is larger. At short times one observes again the presence of
strong finite-size corrections but one can nevertheless appreciate the predicted
scaling behavior $\sim t^{0.72}$ as the system size is increased.

\section{The bulk of the Lyapunov spectrum: Numerical Results}
In this section, we investigate numerically the scaling behavior of $D_{ii}$ for
$i>1$, including the intermediate time regime before saturation, by revisiting
the chain of H\'enon maps and studying the Lorenz-96 model.

It is worth recalling that, in the limit of large system-sizes, the LEs depend
on the integrated density $\rho=(i-0.5)/L$ (in this section we limit ourselves
to studying one-dimensional systems). As a consequence, a meaningful comparison
of LVs for different system sizes must be made by selecting the index $i$ which
corresponds to the same density $\rho$. Since $i$ is, by definition, an integer
variable in the following we interpolate between the two nearest integers that
correspond to the given $\rho$-value.

\subsection{Chain of H\'enon maps}
\begin{figure}
 \centerline{\includegraphics*[width=80mm]{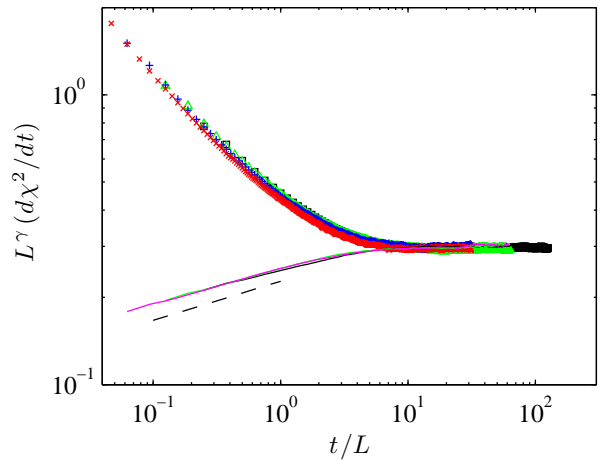}}
\caption{(Color online) Scaled FTLE fluctuations in a chain of H\'enon maps \eqref{henon}, for
$\rho=0.25$. The exponent $\gamma$ is set equal to $\gamma=0.865$. The solid
curves (which correspond to $L=40$, 80, and 160) have been obtained for the
0-norm. The various symbols (squares, triangles, pluses, and crosses
correspond to $L=40$, 80, 160 and 320, respectively) are obtained by using the
Euclidean norm of the Gram-Schmidt LVs. The dashed line corresponds to a power
law growth $(t/L^z)^\nu$ with $\nu = 0.135$ and $z=1$. }
\label{antonio}
\end{figure}

In Fig.~\ref{antonio} we plot the results of simulations performed with the
chain of H\'enon maps for $\rho=0.25$. This $\rho$-value is (i) sufficiently
distant from the singularity at $\rho = 0$  to avoid crossover problems, and
(ii) small enough to be computationally achievable in large
systems~\cite{foot2}. We monitor the evolution of $d\chi^2/dt$ rather than
$\chi^2/t$. The two quantities would be equally valid as both obey the same
scaling relation~\eqref{eq:scal}, but we prefer to use the former one, since it
converges faster (i.e. for smaller values of $t$) to the asymptotic value. The
solid curves in Fig.~\ref{antonio} correspond to simulations performed with the
0-norm for different system sizes. Altogether, the good data collapse confirms
our scaling analysis, with $\gamma\simeq0.865$ (close to the numerical value
0.85 measured in \cite{kuptsov11}) and $z=1$ (as determined from direct LV
studies \cite{szendro07,pazo08}). Moreover, the initial growth agrees with the
theoretical prediction (see the dashed line, whose slope is $\nu =
(d-\gamma)/z\simeq0.135$).

For comparison, in Fig.~\ref{antonio} we plot also the results obtained by
computing the FTLE fluctuations obtained from the standard Gram-Schmidt
orthogonalization procedure (see the symbols). They do not only correspond to a
different way of computing the LEs, but also to a different norm (namely, the
Euclidean or 2-norm). One can see that the asymptotic value of the diffusion
coefficient fully agrees with the previous results: this is consistent with the
expectations that long-time LEs are independent of the norm adopted and this
extends to their fluctuations too. The shape of the corresponding dynamic
scaling-function is, however, very different for both metrics, as expected. In
fact, the LEs exhibit a sub-diffusive transient rather than super-diffusive
behavior if the Euclidean metric is used.

\begin{figure}[t]
 \centerline{\includegraphics*[width=80mm]{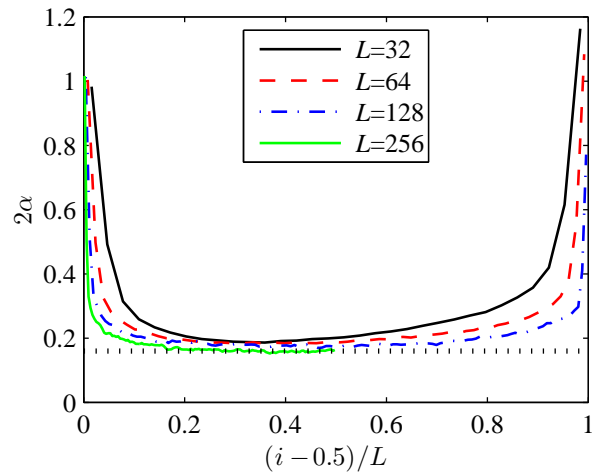}}
\caption{(Color online) Estimation of $2\alpha$ vs.~$\rho$ for the LVs in a chain of H\'enon
maps obtained from the structure factors of the LVs by linear fitting (in
log-log scale) of the 3 smallest wavenumbers. The values of $i$ used in the
$x$-axis correspond to the positive LEs, save for $L=256$ in which case only the
first 128 LEs have been considered due to computational limitations. The dotted
line indicates $2\alpha=0.16$. }
\label{fig_alpha_henon}
\end{figure}

Besides estimating the exponent $\gamma$, we have directly determined $2\alpha$
from the covariant LVs for $\rho$-values below 1, in order to test the validity
of our main relation \eqref{main}. The best estimation of $\alpha$ is typically
obtained from the structure factor (power spectral density) of $h$, which
follows a power-law decay due to its self-affine character,
\begin{equation}
{\cal S}(k) \equiv \lim_{t\to\infty}\langle \hat h(k,t) \hat h(-k,t)\rangle 
\sim k^{-(2\alpha+1)}\, ,
\label{fourier}
\end{equation}
where $\hat h$ is the Fourier transform. A general representation is portrayed
in Fig.~\ref{fig_alpha_henon}, where the effective value of $2\alpha$ is plotted
versus $\rho$ for different system sizes. In the bulk ({\it i.e.}~for
$0<\rho<1$), the data reveals a clear tendency to flatten towards $2\alpha
\approx 0.16$ for increasing the system-size. This $\alpha$-value corresponds to
$\gamma \approx 0.84$, to be compared with the direct estimate $\gamma=0.865$.
This agreement, besides validating relation \eqref{main}, hints at a possible
universal behavior of the LV structure in the bulk of the spectrum. A careful
numerical analysis (analogous to that described in Fig.~\ref{antonio}) for
$\rho=0.75$ (data not shown) further confirms that $\gamma$ is independent of
$\rho$. 

Appreciable deviations from the basal value $2\alpha \approx 0.16$ are clearly
visible in Fig.~\ref{fig_alpha_henon} in the vicinity of the two extreme values
$\rho \to 0$ and $\rho \to 1$. For $\rho\ll 1$, we know that the first LV
follows KPZ scaling, $2\alpha=1$, and the data must show a cross-over towards
such a different scaling, when $\rho=0.5/L\approx 0$ is approached. A similar
behavior is found for $\rho=1$. Note that this value does not 
correspond to the smallest
LE (which is obtained for $\rho=2$, as there are $2L$ exponents in a system of
size $L$), but rather to the edge of the band of positive LEs.
In fact, for these parameter values there exists a gap between the positive and
negative bands of LEs: the singularity for $\rho=1$ thus reinforces the idea that band
edges are characterized by a quantitatively different behavior, although here we do
not see a direct way to map the evolution onto, e.g., KPZ dynamics.

\subsection{Lorenz-96 model}
\begin{figure}
 \centerline{\includegraphics*[width=80mm]{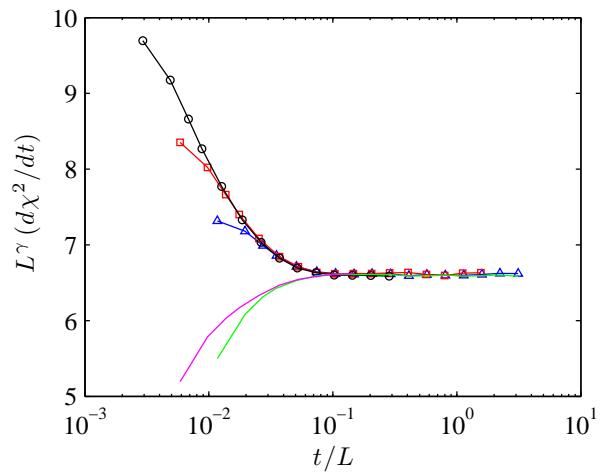}}
\caption{(Color online) Scaled FTLE fluctuations in the Lorenz-96 model \eqref{l96}, for
$\rho=0.1$. The exponent $\gamma$ is set equal to $\gamma=0.897$. The solid
curves, corresponding to $L=128$ and 256 have been obtained from the covariant
LVs using the 0-norm. The various symbols (triangles, squares, and circles
correspond to $L=128$, 256, and 512, respectively) are obtained by using the
Euclidean norm during the forward iteration of the Gram-Schmidt LVs. }
\label{fig_rho01}
\end{figure}

The most intriguing message that arises from the study of the H\'enon maps is
the possibly universal scaling behavior of the bulk LVs. Given the relevance of
such an observation, we  have studied also the Lorenz-96 model \eqref{l96}.
Being a continuous-time system, simulations are heavier than in the previous
case and for this reason we have been able to carry out extensive simulations
only for $\rho=0.1$ which is nevertheless far enough from $\rho=0$ to draw
meaningful conclusions. The results for $L$ up to 256 for the 0-norm and
covariant LVs, and up to 512 for the Gram-Schmidt LVs, are plotted in
Fig.~\ref{fig_rho01}. The good data collapse confirms that the dynamic exponent
is $z=1$ as in the previous model. As for $\gamma$, we find a slightly different
value, namely $\gamma=0.897$ (to be compared with $\gamma=0.865$). So far it is
not possible to determine whether this difference is significative or just due
to strong model-dependent finite-size effects hiding a universal
system-independent value. However, the closeness of both numbers suggests that
$\gamma$ is universal in the bulk Lyapunov spectrum.

\begin{figure}[t]
 \centerline{\includegraphics*[width=80mm]{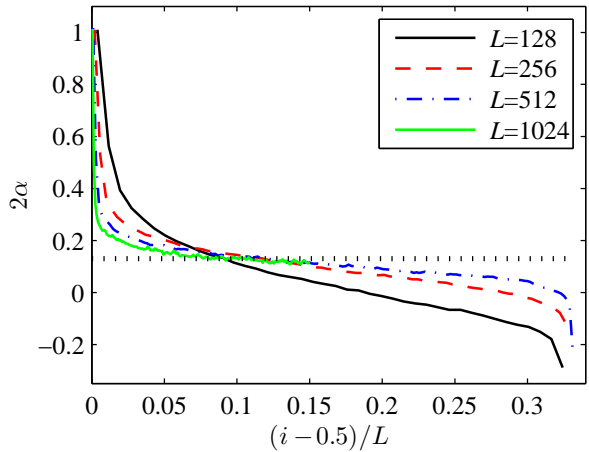}}
\caption{(Color online) Estimation of $2\alpha$ vs.~$\rho$ for the Lorenz-96 model obtained
from the structure factors of the LVs by linear fitting (in log-log scale) of
the 3 smallest wavenumbers. The values of $i$ used in the $x$-axis correspond to
the positive LEs, save for $L=1024$ in which case only the 152 largest LEs have
been considered due to computational limitations. The dotted line indicates
$2\alpha=0.13$.}
\label{fig_alpha_l96}
\end{figure}

We have also determined the values of $\alpha$ from the structure factors ${\cal
S}(k)$ of the LV-surfaces. The values of $2\alpha$ for the positive LEs are
plotted in Fig.~\ref{fig_alpha_l96}. As shown above for the chain of H\'enon
maps, here the curve also becomes increasingly flat as the system size grows
except for the points close to $\rho=0$ and $\rho\simeq0.33$: (i) $\rho=0$
corresponds to the first LE, where we know that $2\alpha=1$; (ii)
$\rho\simeq0.33$ corresponds to
the vanishing Lyapunov exponent \cite{pazo08}, so that ${\cal S}(k)\sim k^0$, 
{\it i.e.} $2\alpha=-1$. These results suggest the existence of
a common $\alpha$ for the bulk in the thermodynamic limit. Our best estimation
is $2\alpha\simeq0.13$, not far from the estimation $0.15$ in \cite{pazo08}.
Moreover, assuming $z=1$, the value of $\gamma$ expected via Eq.~\eqref{main} is
0.87, which is relatively close to the value $0.897$ observed in
Fig.~\ref{fig_rho01}.

\section{Conclusions and open problems}
Altogether, in this paper we have shown that the analogy between roughening
phenomena and LV dynamics in spatially extended systems is rather fruitful in
that it allows relating the scaling behavior of the LE diffusion coefficients
with the roughness exponents of the corresponding vectors. Numerical simulations
of various models support the general scaling relation~\eqref{eq:scal} and of
its asymptotic behavior~\eqref{d} and~\eqref{eq:delta}. With reference to the
first LV, our analysis confirms the validity of the relationship with the KPZ
equation in dissipative systems.

More intriguing is the question of the scaling behavior for the
bulk of the Lyapunov spectrum. Our studies of H\'enon maps and of the Lorenz-96
model reveal that in both cases, independently of $\rho$, $z=1$ and
$\alpha\approx 0.07$. Nonetheless, small but not-so-negligible differences
for the latter exponent are observed. 

Two questions are, therefore, still open in the case of the fluctuations of the
bulk LEs: (i) whether $\alpha$ is strictly larger than zero ($\gamma<1$) in the
thermodynamic limit; (ii) the very existence of a single universality class. The
$\gamma<1$ issue arises from the comparison with the cross-correlations of
different LEs-- namely, the scaling of the off-diagonal terms $D_{ij}$ with $i
\neq j$. In Ref.~\cite{kuptsov11} such terms have been found to scale as $1/L$,
{\it i.e.}~their $\gamma$-value is 1. This finding has been interpreted as the
evidence of an extensive behavior of the large deviation function (which is
proportional to the inverse of $\bf D$). On the one hand, it is strange that the
diagonal terms scaling differs from that of the off-diagonal ones. On the other
hand, the very fact that $\gamma<1$ implies that $\alpha>0$. This, in turn,
indicates a strong localization of the covariant LV, a property that has been
observed in different models by using different
methods~\cite{szendro07,ginelli07}. On the basis of the results derived in this
paper for the H\'enon maps (that are more reliable than those for the Lorenz-96
model) we feel confident in stating that $\gamma<1$. In order to draw firmer
conclusions, however, we believe that it is necessary to make some substantial
progress on the theoretical side by either identifying a minimal stochastic
model of the LV dynamics for $i\gg 1$ (such as the KPZ equation for the first
vector), or by resorting to new results within the field of random matrices.

\acknowledgments
DP acknowledges support from Ministerio de Econom\'ia y Competitividad (Spain)
under a Ram\'on y Cajal fellowship, and from Cantabria International Campus.
We acknowledge financial support from
MICINN (Spain) through project No. FIS2009-12964-C05-05.


\begin{thebibliography}{10}
\bibitem{foot1}Strictly speaking, the average
over different trajectories is not needed in the infinite-time limit. We prefer,
however, to keep the angular-brackets since almost everywhere in this paper,
we refer to finite-time quantities.
\bibitem{foot2} In fact, the  number of
operations needed to compute the $i$-th LE is of the order of $i^2 N R$, where
$N$ is the phase-space dimensionality --- here $N=2L$ --- and $R$ the number of
orthogonalizations to be performed (at variance with the computation of the
surface square width $W^2$, a much larger statistics is here needed to determine
the LE fluctuations $\chi^2$).
\end{thebibliography}
\end{document}